\pdfoutput=1

\documentclass{llncs}
\usepackage{float}
\usepackage{eurosym}
\usepackage{subfigure}
\usepackage{stfloats}
\usepackage{amsfonts}
\usepackage{amsmath}
\usepackage[pdftex]{graphicx}
\usepackage[utf8]{inputenc}
\usepackage{booktabs}
\usepackage{color}
\usepackage{array}
\usepackage[table]{xcolor}
\usepackage{epstopdf}
\begin{document}
\newcolumntype{C}[1]{>{\centering\arraybackslash}p{#1}}

\title{Choosing the right home location definition method for the given dataset}

\author{Iva Bojic, Emanuele Massaro, Alexander Belyi, \\ Stanislav Sobolevsky, Carlo Ratti}

\institute{Senseable City Lab, Massachusetts Institute of Technology, Cambridge, MA, US \\ SMART Centre, Singapore \\
Email: ivabojic@mit.edu, emassaro@mit.edu, alex.bely@smart.mit.edu, stanly@mit.edu, ratti@mit.edu}

\maketitle

\begin{abstract}
Ever since first mobile phones equipped with Global Position System (GPS) came to the market, knowing the exact user location has become a holy grail of almost every service that lives in the digital world. Starting with the idea of location based services, nowadays it is not only important to know where users are in real time, but also to be able predict where they will be in future. Moreover, it is not enough to know user location in form of latitude longitude coordinates provided by GPS devices, but also to give a place its meaning (i.e., semantically label it), in particular detecting the most probable home location for the given user. The aim of this paper is to provide novel insights on differences among the ways how different types of human digital trails represent the actual mobility patterns and therefore the differences between the approaches interpreting those trails for inferring said patterns. Namely, with the emergence of different digital sources that provide information about user mobility, it is of vital importance to fully understand that not all of them capture exactly the same picture. With that being said, in this paper we start from an example showing how human mobility patterns described by means of radius of gyration are different for Flickr social network and dataset of bank card transactions. Rather than capturing human movements closer to their homes, Flickr more often reveals people travel mode. Consequently, home location inferring methods used in both cases cannot be the same. We consider several methods for home location definition known from the literature and demonstrate that although for bank card transactions they provide highly consistent results, home location definition detection methods applied to Flickr dataset happen to be way more sensitive to the method selected, stressing the paramount importance of adjusting the method to the specific dataset being used.
\end{abstract}

\section{Introduction}

A high portion of our daily lives is also happening in digital world. People can express themselves using microblogging platforms such as Twitter; stay in a contact with their family and friends using Online Social Networks (OSNs) such as Facebook; check in at different locations using Location Based Social Networks (LBSN) such as Foursquare or share their photos from vacations using Flickr. 

All of those online platforms offer people to create their own digital content that is very often public and consequently to a certain extent available not only for scholars in their scientific studies, but also for city policy makers on urban scales or for marketing purposes world-wise. With mobile phones equipped with a highly precise Global Position System (GPS) modules becoming omnipresent and the increase in sharing our location within different mobile apps, world has become richer for an enormous source of data on human mobility. Digital data not only does play a crucial role in research on human behavior and human mobility \cite{calabrese2011estimating,gonzalez2008Understanding,hoteit2014estimating,kung2014exploring}, but also opens new unprecedented opportunities for supporting planning decisions, such as regional delineation \cite{ratti2010redrawing,sobolevsky2014money,sobolevsky2013delineating} or land use classification \cite{grauwin2015towards,pei2014new,sobolevsky2015bbva,sobolevsky2015cities,sobolevsky2014mining}, as well as for instance paving a way towards a smarter urban transportation optimization \cite{santi2014quantifying}. 

Although the amount of publicly available data is rising, a lot of information is still missing and has to be inferred from the data that is shared. In that sense, a lot of related work is dedicated towards building different \textit{location to profile} frameworks using spatial, temporal and location knowledge for example to infer demographic attributes of users sharing their location \cite{Zhong2015You} or labeling important places in somebody's life \cite{Krumm2013Placer}. In this paper we focus on predicting user home locations for different online platforms, as well as for everyday traces that people leave in digital world outside OSNs (e.g., bank card transactions, call records). Namely, the exact locations where people live or work are very often ambiguous and are missing along with other pieces of human mobility puzzle. As mentioned in \cite{Li2012Towards} only 16\% of Twitter users reported their city level location. Moreover, it was shown that 34\% people reported invalid information or they even put sarcastic comments \cite{Hecht2011Tweets}. Inferring home location is just one special case of the process in which semantic place labels such as home, work or school are given to geographic locations where people spend their time. Information on home location is not only important for investigating mobility patterns, but also for delivering localized news, recommending friends or serving targeted ads \cite{Li2012Multiple}. It can be also helpful when designing more personalized urban environments (e.g., pollution management, transportation systems) or modeling outbreaks and disease propagation \cite{Zheng2014Inferring}.

The contribution of our work is twofold. First, we show how the basic mobility patterns quantified by means of the radius of gyration are different in cases of digital trails of bank card transactions and Flickr. Second, we show how different nature of the datasets affects the applicability of various methods for inferring user home locations by comparing the outcomes of five simple home definition methods applied to the two aforementioned datasets. The rest of the paper is structured as follows. Section \ref{rw} summarizes related work on methods for determining home locations, while Section \ref{dataset} introduces two datasets used in this work: Flickr dataset and dataset of bank card transactions. In Section \ref{rog} we present results of radius of gyration applied to the two aforementioned datasets and then in Section \ref{hd} we compare results of five different home definition methods for those two datasets. Finally, Section \ref{con} concludes the paper and gives guidelines for future work.

\section{Related work}
\label{rw}

Related work can be mostly divided into two lines of work: studies that only focus on finding suitable algorithms for predicting where people live or studies that focus on topics for which knowing home locations is the prerequisite. In the latter group of studies scholars mostly use simple methods for determining where people live such as maximal number of geolocalized tweets/photographs/check-ins, which might work well for some of the datasets, but not necessarily for all of them. In that sense, the context of the dataset is often not really considered. With this work we want to raise the concern that before using simplified home definitions, one must be aware of the said context, as otherwise using an inappropriate method can cause uncontrollable errors in inferring home locations.

In this paper we consider three different types of OSNs: microblogging platforms such as Twitter, LBSNs such as Foursquare and photo-sharing sites such as Flickr. In addition to them, we also distinguish digital traces that people leave and that do not have a direct social component such as Detail Call Records (DCRs) or bank card transactions. As mentioned before, the most commonly used method to infer where people live is to assume that it is the location from which they sent the maximal number of tweets in case of studies on Twitter \cite{hawelka2014geo,Sobolevsky2015Scaling}, where they had the highest cell phone communication activity in case of DCR datasets \cite{Onnela2011Geographic} or where they did the maximal number of check-ins in case of LBSNs \cite{Noulas2012Random}. In addition to this \textit{max*} method, in some cases together with the max number of '*', time also plays an important role when inferring home locations for Flickr \cite{Bojic2015Sublinear,Paldino2015Urban}, Gowalla and Brightkite \cite{Cho2011Friendship} or DCRs \cite{Alexander2015Origin,kung2014exploring}.

\subsection{Estimating user locations: content approach}

Even before trying to extract information about user location using content approach, one must face with a challenge of detecting the right language for the given text \cite{Graham2014World}. Once when the right language is detected, the next step is to extract information about user location from the text which is usually done in two ways -- searching for geographic hints or building probabilistic language models for a specific location. In the former case words in the text are compared against a specialized external knowledge base, while in the latter case messages posted from the same location are clustered based on the word usage in them.

The problem when matching locations shared in the user generated text with information stored in an external dataset (e.g., gazetteer) is the ambiguity for examples for cities that do not have country/state associated with them and their names are not unique or when non-location related words are matched with cities/towns. Possible solutions for dealing with this are to: 1) use location priors, 2) search for disambigitors or 3) apply spatial minimality method \cite{Serdyukov2009Placing}. The first method proposes to infer user location following the simple rule that places with larger populations or places that are more frequently mentioned in the text are more likely to be candidates, while the second and third one propose to use a list of disambigitors for every place or to calculate the minimal bounding rectangle containing all of them and then again choosing the most likely one.

When using probabilistic language models, two approaches can be distinguished -- building a language model for a city estimated from tweet messages \cite{eisenstein2010latent,kinsella2011m} or calculating the city distribution on the use of each word \cite{chang2012phillies,Cheng2010You}. The first approach assumes that users living in the same city use similar language (i.e., language usage variations over cities), while the second one concentrates on calculating spatial word usage. It was shown that selected set of words (i.e., local words) can be used as a stronger predictor of a particular location \cite{Cheng2010You}. The selection process of local words can be done either using a supervised classification method like in \cite{Cheng2010You} or unsupervised one like in \cite{chang2012phillies}. Both approaches achieved the accuracy of around 50\% predicting user home locations within 100 miles.

Although text analysis is the most commonly chosen method for estimating user home locations using content approach because text is present in almost all human digital traces, scholars have also used other content sources (e.g., photographs, audio, video) to infer where people live. In that sense in \cite{Zheng2014Inferring} and \cite{zheng2015towards} authors used visual features of photographs to distinguish between "home" and "non-home" photographs. As claimed by authors, unlike photo tags and descriptions, which do not have to be available, visual content is always available for every photograph. It was reported that the proposed method for home prediction achieved an accuracy of 71\% with a 70.7-meter error distance.

\subsection{Estimating user locations: social and historical tie approach}

Even when users think they are cautious enough not to reveal not only their locations while using OSNs, but also where they live, their less cautious friends can implicitly \textit{fill out} these gaps for them. Namely, users with known locations can be treated as noisy location sensors of their more privacy-aware friends, as it was observed that likelihood of friendship with a person is decreasing with distance \cite{backstrom2010find}. Moreover, the total number of friends tends to decrease as distance increases. As a result, it was shown that over half of users in Twitter network have one friend that can be used to predict their location within 4 km. This is not surprising given that our social options are not endless, but in fact constrained in sense that it takes time, energy and money to maintain them. However, the interesting finding was that our online activity is still influenced by locality from the physical world \cite{Jurgens2013Thats}.

One of the first papers in the field was from Backstrom in 2010 who predicted the location of Facebook US-based users using location information of their friends \cite{backstrom2010find}. His results showed that almost 70\% users with 16 and more friends can be placed within 25 miles of their actual home. In later years, by defining a user home location in Twitter network as the place where most of his/her activities happen and assuming that 1) a user is likely to follow and be followed by users who live close to him/her and 2) in his/her tweets he/she may mention some "venues" which may indicate his/her location, in \cite{Li2012Towards} scholars presented their method for which they showed improvement of 13\% when compared to state-of-the-art methods. Additionally, Chen et al.\ showed that these results can even be improved when adding information how strongly users are connected and when trusting more friends with whom users interact more frequently \cite{chen2014tie}.

\section{Dataset}
\label{dataset}

In this study we use two different datasets -- a complete set of bank card transactions recorded by Banco Bilbao Vizcaya Argentaria (BBVA) during 2011 in the whole Spain\footnote{Although the raw dataset is protected by a non-disclosure agreement and is not publicly available, certain aggregated data may be shared upon a request and for the purpose of findings validation.} and Flickr dataset for the whole world created by merging two publicly available Flickr datasets \cite{flickr1,flickr2,thomee2015new}. In BBVA dataset, transactions were performed by two groups of bank card users. The first one consists of the bank direct customers, residents of Spain, who hold a debit or credit card issued by BBVA. In the considered time period, the total number of active customers was more than 4 million, altogether they executed more than 175 million transactions in over 1.2 million points of sale, spending over 10 billion euros. The second group of card users includes over 34 million foreign customers of all other banks abroad coming from 175 countries, who made purchases at one of the approximately 300 thousand BBVA card terminals. In total, they executed another 166 million transactions, spending over 5 billion euro. Flickr dataset used in our study has more than 1.25 million users who took more than 130 million geo-tagged photographs/videos in 247 countries around the world within a ten-year time window period (i.e., from 2005 and until 2014).

In BBVA dataset for each transaction we know: date and time when it happen associated with the anonymous customer and business IDs and amount of money that was spent. IDs of customers and businesses are connected with certain demographic characteristics (e.g., age group and gender) and location (i.e., latitude and longitude coordinates) where people live or where businesses are located. Moreover, business IDs are associated with business categories they belonged to. In total there are 76 categories such as restaurants, gas stations or supermarkets. In Flickr dataset each photograph/video has its own ID associated with unique user ID, time stamp, and geo coordinates. For some photographs/videos we have also additional information such as: title, description, user tags, machine tags, photograph/video page download URLs, license name and URLs. However, unlike in BBVA dataset, Flickr dataset does not contain any additional information about users that took photographs/videos.

\begin{figure}[t!]
  \centering
      \includegraphics[width=0.94\textwidth,trim={0 0 6cm 0}]{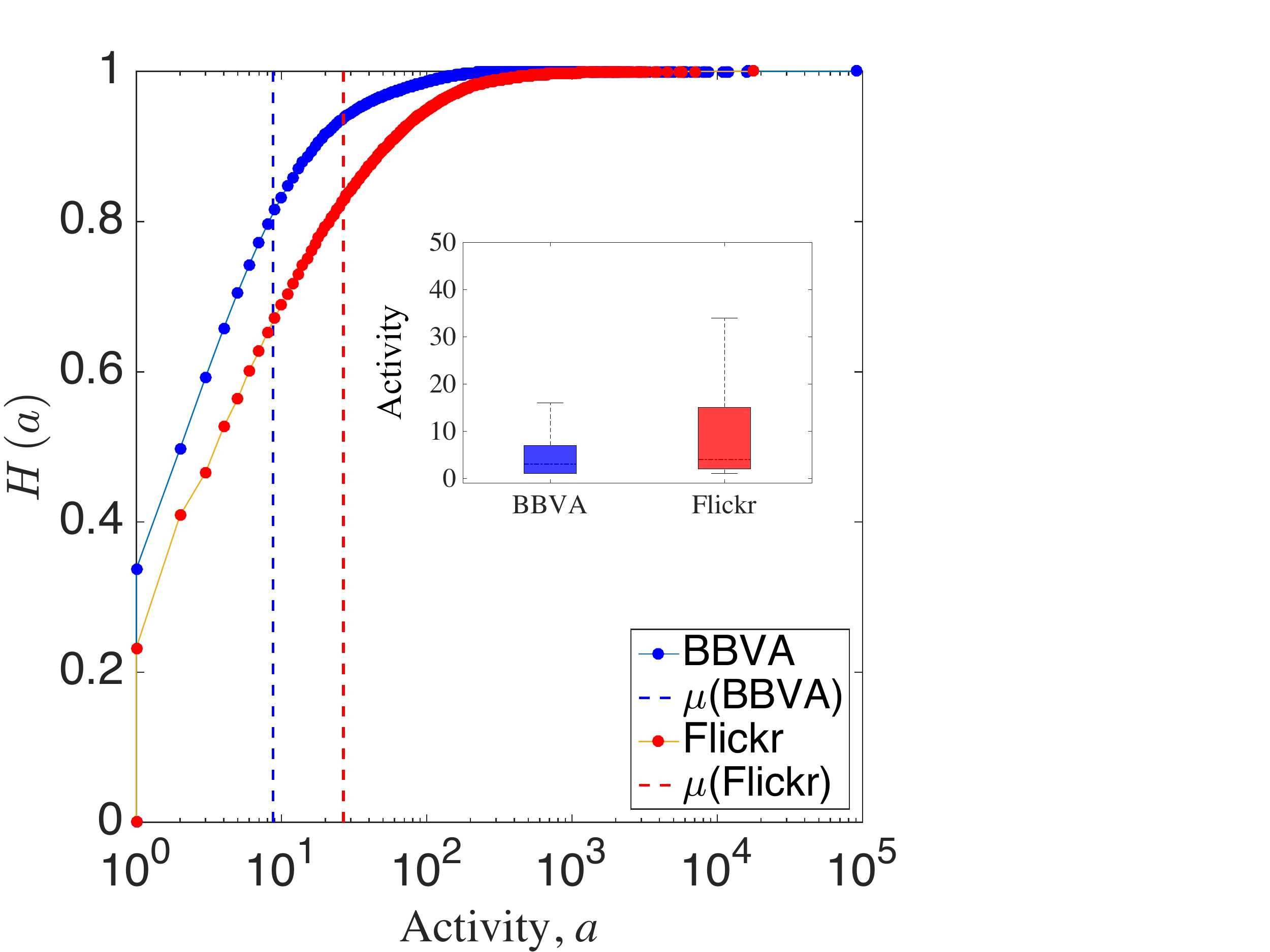}
  \caption{(\emph{xaxis log}) We report the empirical cumulative density function ($H(a)$) of the users' activity for both the datasets and the statistical boxplots in the top right. The "central box" representing the central $50\%$ of the data. Its lower and upper boundary lines are at the $25\%$-$75\%$ quantile of the data. The central line indicates the median of the data. The average activity in the BBVA dataset (i.e., average number of transactions) is $\hat{a}^{BBVA}\sim8.75$ while the Flickr users made in average $\hat{a}^{Flickr}\sim27$ photographs/videos while medians, represented by horizontal dotted lines in the boxplots, are respectively $3$ and $4$.}
 \label{fig:datasets}
\end{figure}

Data preprocessing for both datasets is implemented as a two-step process: first we prune the data and then we filter users/customers that are not active enough, i.e., for whom we do not have enough information. In the case of BBVA dataset pruning is done in such a way that we omit all customers for whom we do not have records of their exact home location, and all photographs/videos in Flickr dataset for which reverse geo-codding did not return any results or was inconclusive (i.e., we could not determine country where they were taken). \figurename~\ref{fig:datasets} shows resident and foreign customer activity in BBVA dataset and activity of all users that took at least one photograph/video in Spain during the observed time window of ten years in Flickr dataset. This was the data that was used to calculate radius of gyration which results was shown in Section \ref{rog}.

Unlike in the case of BBVA dataset, for Flickr dataset we do not have ground truth that would tell us where people live. In order to compare resident activity from BBVA dataset with resident activity from Flickr dataset, we had to put a firm threshold on Flickr user activity to make sure that we include only Spanish users. In that sense, when determining home location, we first choose users that took at least one photograph/video in Spain (105\,918 users in total) and then take their activity all around the world which brings us from 5\,758\,938 photographs/videos that those users took only in Spain to 21\,134\,113 photographs/videos they took worldwide. 

After having the complete data for Flickr dataset, we run five home location definition methods, which are explained in Section \ref{hd}, to first determine their home country. We consider that a particular user lives in Spain only if the result was consistent over all five methods, which leaves us in the end with 19\,336 users for whom Spain is their home country. Finally, when determining home locations on 52 Spanish provinces level for customers/users from both datasets, we include only those who made more than the average number of photographs/videos/transactions. Table \ref{table:datasets} shows the total numbers of customers/users/photographs/videos/transactions and how those numbers changed when applying filters and methods for pruning the data.

\begin{table}[h!]
\centering
\caption{Datasets statistics.}
\begin{tabular}{l| C{2cm} | C{2cm}}
                                         			& BBVA          &  Flickr       \\\hline
Number of users/customers for radius     			& 39.3~mln    &  105.4~k      \\
Number of transactions/photos for radius 			& 341.4~mln  &  5.2~mln    \\
Number of users/customers for home definition      	& 1.2~mln     &  2.5~k        \\
Number of transactions/photos for home definition 	& 141.7~mln   &  2.4~mln
\end{tabular}
\label{table:datasets}
\end{table}

\vspace{-20pt}

\section{Radius of gyration}
\label{rog}

It is undoubtedly that on average people spend most of their time at home as it was also shown by numerous studies (e.g., \cite{Pontes2012We}). However, the question that we pose here is if different datasets show that pattern in the same way. Our assumption is that because of the different nature how BBVA and Flickr datasets were created, we will demonstrate that they reveal different human mobility patterns and consequently that we cannot apply to them the same methods for predicting home locations. In this section we thus present results of radius of gyration applied to BBVA and Flickr datasets where the linear size occupied by each user's trajectory up to time $t$ is defined as \cite{gonzalez2008Understanding}:

\begin{equation}
r^a_g(t) = \sqrt{\frac{1}{n_c^a(t)} \sum_{i=1}^{n^{a}_c}  \left(  \vec{r_i^a} - \vec{r^a_{cm}}  \right)^2 }
\end{equation}

\noindent
where $\vec{r_i^a}$ represents the $i=1,2,....,n^a_c(t)$ position visited by user $a$, while $r^a_{cm}=1/n_a^c(t)\sum_{i=1}^{n^a_c} \vec{r_i^a}$ is the center of mass of the trajectory.

In general, radius of gyration refers to the distribution of the components of an object around an axis. In particular, we use radius of gyration in this paper as it has already been proven that it can be used as a good proxy for human mobility. Unlike for example in the case of average travel distance, the radius of gyration is smaller for a user who travels in a comparatively confined space even though he/she covers a large distance, but is larger when someone travels with small steps but in a fixed direction or in a large circle. Consequently, it captures exactly the same pattern that we are interested to compare for our two datasets -- time dependent dissipation of user movements.

\begin{figure}[b!]
\vspace{-20pt}
  \centering
      \includegraphics[width=1\textwidth,trim={0 0 6cm 0}]{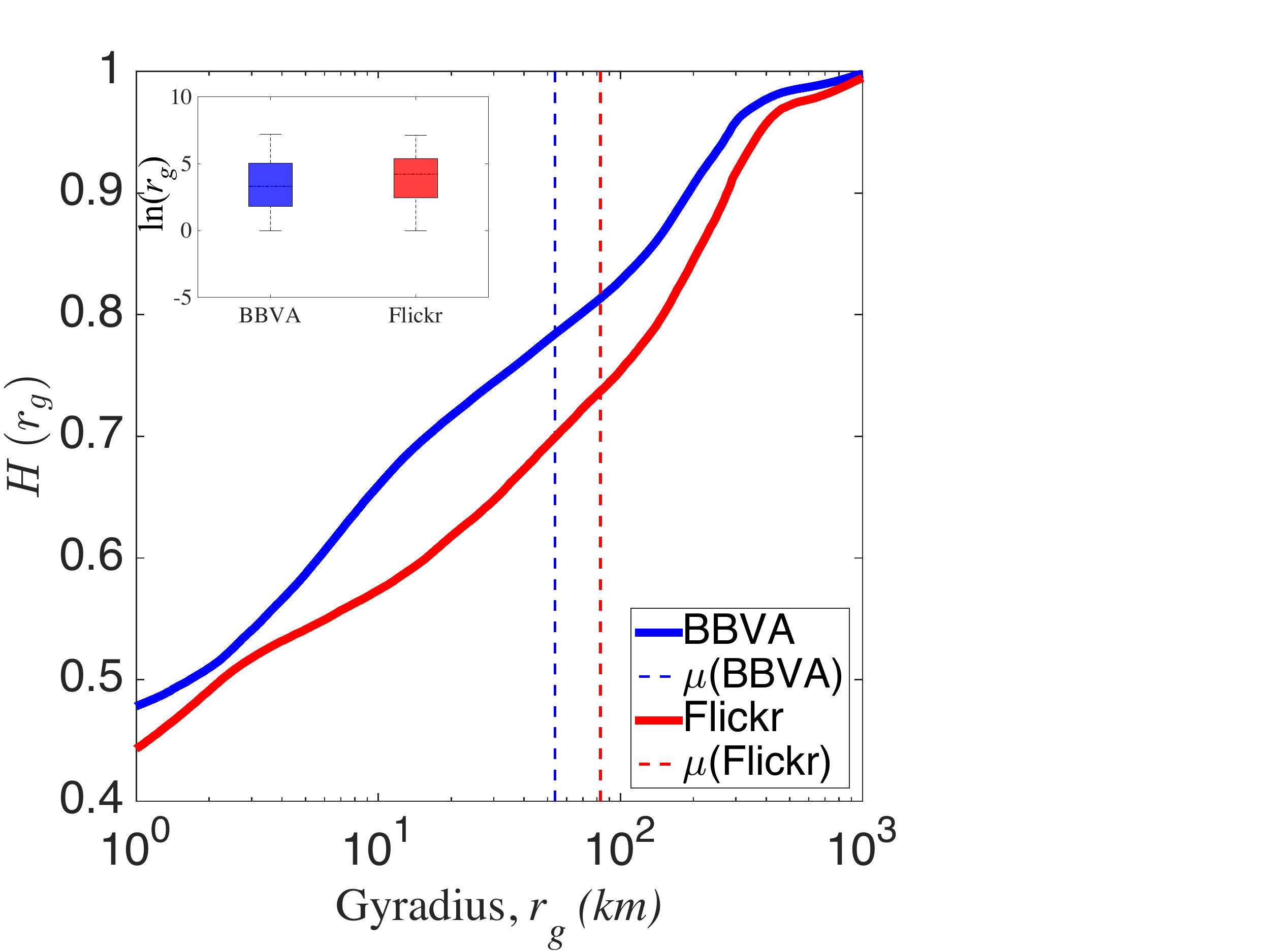}
  \caption{We report the empirical cumulative density function ($H(r_g)$) for both the datasets (the boxplots of the natural logarithm of the radius of gyration on the top left). The average radius of gyration (i.e., dotted lines in the figure) for the BBVA and the Flickr dataset are respectively $\hat{r_g}^{BBVA}\sim 54$ $km$ and $\hat{r_g}^{Flickr}\sim 83$ $km$, while medians are respectively $1.67$ $km$ and $2.25$ $km$.}
\label{fig:distrrad}
\end{figure}

The distribution of the radius of gyration, which is shown in \figurename~\ref{fig:distrrad}, can reveal interesting human mobility patterns captured in Flickr and BBVA datasets. The most interesting result is that the radius of gyration is higher in Flickr than in BBVA dataset, with the averages $83$ $km$ and $54$ $km$ respectively. Moreover, from Flickr dataset it can be observed that $30\%$ of users travel more than $100$ $km$ compared to only $20\%$ in BBVA scenario. Although both datasets include activity of residents and tourists, differences between their behavior are more emphasized in Flickr dataset supporting our initial assumption. Namely, the travel pattern observed from Flickr can be explained that it is more often used when people travel as it has already been shown that tourists generally have a higher and sparser travel activity during the visit to a new country \cite{hawelka2014geo}.

\begin{figure}
\centering
\subfigure[]{\includegraphics[width=0.85\textwidth]{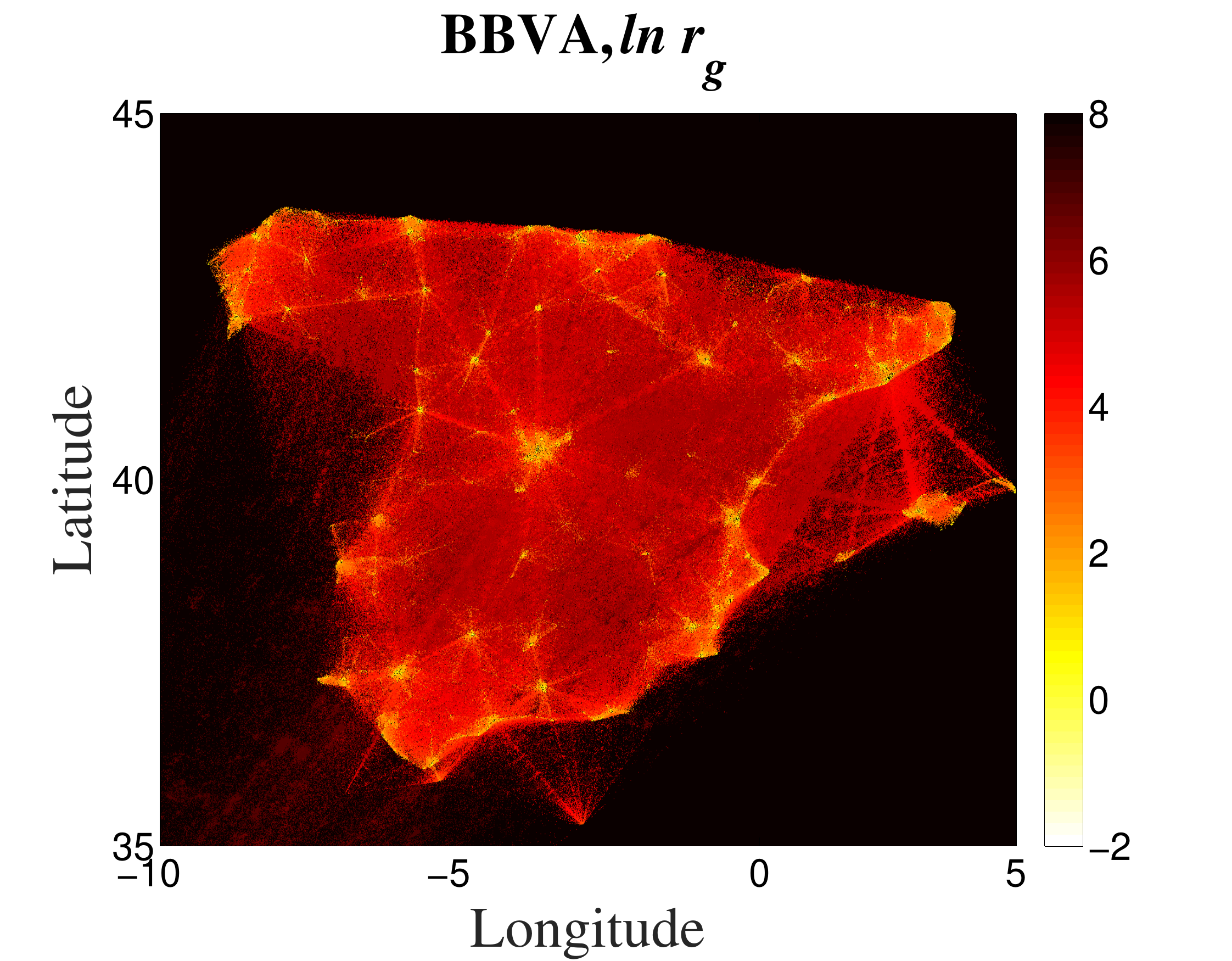}}
\subfigure[]{\includegraphics[width=0.85\textwidth]{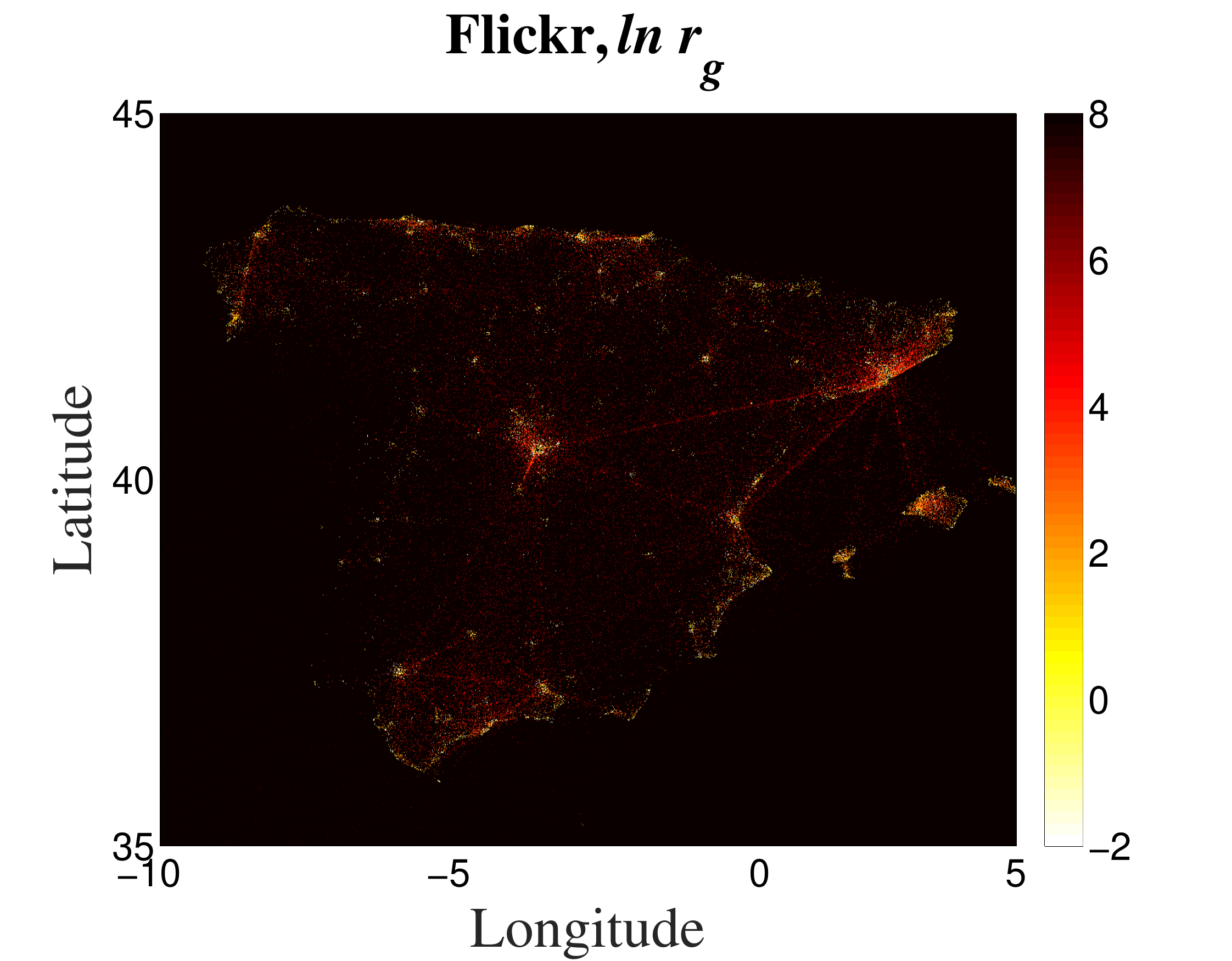}}
  \caption{Radius of gyration ($ln\left(r_g\right)$) for the users in the (a) BBVA and (b) Flickr datasets. Each point corresponds to the center of mass of the users, while the color represents the value of the radius of gyration.}
\label{fig:bbvarg}
\end{figure}

In~\figurename~\ref{fig:bbvarg}, each point corresponds to the coordinate of the center of mass of $r_{cm}$ for each customer/user from both dataset, while the color highlights its (natural logarithmic) value. From these two maps we can observe the fact that people who have their center of mass in big cities have a smaller radius of gyration than people who live in suburban areas meaning that in general people who lives in urban areas travel less than people who live in rural areas or small villages \cite{pateman2011rural}. This result makes a perfect sense because most of the day life services (e.g., from work to shopping, from free time to education) are concentrated in the cities and people who live outside tend to travel there for most of their activities. 

\section{Home detection methods}
\label{hd}

In this section we apply five very simple home detection methods, which are very often used in related work, to Flickr dataset and dataset of BBVA bank card transaction records. The goal of our paper is not to propose a new method, but to compare how results change in case of different datasets. With this we want to show that one should calculate in for dataset differences when deciding which method to use. The home inferring methods that we are using are the following:

\begin{enumerate}
\item home location is inferred as a place where a user/customer took/made maximal number of photographs/videos/transactions,
\item home location is inferred as a place where user/customer spend the maximal number of active days, where an active day is a day when user/customer took/made at least one photograph/video/transaction,
\item home location is inferred as a place with the maximal timespan between the first and the last photograph/video/transaction,
\item home location is inferred as a place where a user/customer took/made maximal number of photographs/videos/transactions from 7 PM to 7 AM (when users/customers are supposed to be near home)
\item home location is inferred as a place where user/customer spend the maximal number of active days from 7 PM to 7 AM (when users are supposed to be home).
\end{enumerate}

\begin{figure}[h!]
  \centering
      \includegraphics[width=1\textwidth,trim={2cm 2cm 3cm 2cm}]{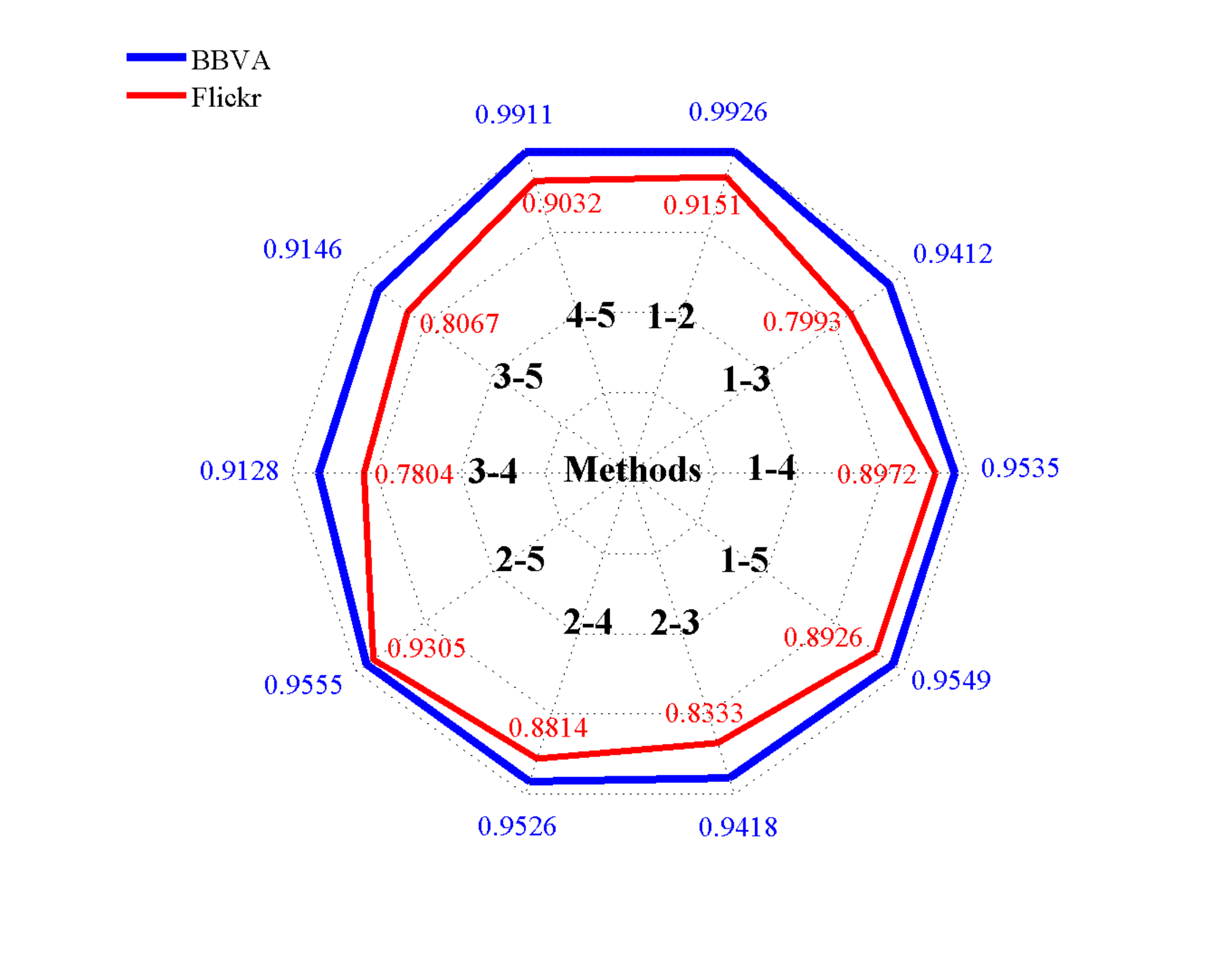}
  \caption{Radar plot showing comparison of results (in terms of pairwise SMC) of 5 different home detection methods for BBVA and Flickr. Radar axes denote different pairs of home definition methods.}
\label{fig:smc_radar}
\end{figure}

Once when calculated home definition methods for all customers/users, for each method and for each dataset we generate a vector $Prov_{method\_x}$ containing province numbers that represent customers/users estimated home locations. For example, $Prov_{method\_3}[i] = 34$ denotes that inferred home province for customer/user $i$ using method 3 is 34. The results of comparison between different home definition methods applied to the same dataset are reported in~\figurename~\ref{fig:smc_radar}. To compare results of two different methods we used Simple Matching Coefficient (SMC). This measure shows how similar two vectors are. In our case we had two vectors of length~$n$: $Prov_{method\_x}$ and $Prov_{method\_y}$ containing province numbers assigned to~$n$ customers/users and we calculate SMC as:
\[
	SMC(method\_x, method\_y) = \frac{\sum_{i=1}^{n}{\delta(Prov_{method\_x}, Prov_{method\_y}})}{n},
\]
where $\delta(x, y)$ equals to~$1$ if~$x=y$ and~$0$ otherwise.

\newpage

The value of SMC basically represents a fraction of results, identical for two selected methods, to the total number of all results and as such can be used to compare how applying of different methods affects the final results. \figurename~\ref{fig:smc_radar} shows that for BBVA dataset differences between comparison of all pairs of five used home detection methods are less than 9\% (worse results are when comparing methods 3 and 5) and can go as high as less than 1\% in case when comparing methods 1 and 2. In Flickr dataset final results of comparison can differ more than 20\% in case of home detection methods 3 and 4 and they are never less than 7\% (e.g., for methods 2 and 5). Moreover, an interesting finding is that based on results of these two datasets we cannot conclude which two methods are the most similar/diverse. Nevertheless, results shown in \figurename~\ref{fig:smc_radar} support our assumption that different datasets are not equally susceptible when applying different home definition methods, providing a proof that it is of vital importance to choose the "correct" method with the respect of the specific dataset being used.

\vspace{15 pt}

\section{Discussion and conclusions}
\label{con}

With the emergence of digital datasets (e.g., bank card transaction records or detail phone records) containing interesting information about human race, scholars can give answers to questions we have never thought could be answered before. However, the available data represents just a certain sample of human activity, and often a very heterogeneous one, which might lead to certain mistakes that can be introduced into the process of discovering new things when not fully understanding used data sources. In this paper we focused on showing how the usage of oversimplified home definition methods can change the final results when semantically labeling most probable places where people live. Although process of inferring people home locations can be a focus of standalone studies, it is in fact very often a (first) part of more complex studies on for example human mobility or personalized services and as such is important for a larger body of related work.

Most of the simple home definition methods rely on the assumption that people spend the maximal amount of their time at home and that it can be inferred in the same way no matter what kind of dataset is used. In order to see if this assumption holds, we chose on purpose two very different digital datasets of Online Social Networks (OSNs) and Banco Bilbao Vizcaya Argentaria (BBVA) bank card transaction records. Namely, if we can show this is true in case of two datasets that are so different, then it will also hold for datasets that are more similar (e.g., two OSNs). To check if there is any difference of how different datasets capture human mobility, we first compared distributions of the radius of gyration and then we applied five different simple home definition methods to the aforementioned datasets.

The distribution of the radius of gyration showed that indeed there are observable differences how and where people use their bank cards and Flickr. Namely, with almost 50\% of larger radius of gyration and 10\% of more people who traveled more than $100$ $km$, Flickr dataset provides another view on human mobility than BBVA dataset. Moreover, BBVA dataset seems to be more robust to different home definition methods as the results differ only from 1\% to 9 \% unlike in case of Flickr that are in range of 10\%--20\%. With this we showed that although for some datasets (such as bank card data) particular choice of a home definition method does not really seem that important, for the other datasets, it might actually affect the results quite a bit. Therefore, the choice of home definition method should be done carefully with respect to the characteristics of the particular dataset being considered.

However, results presented in this paper are only the initial results that pointed towards differences between human mobility patterns captured in two different datasets and should be extended to other datasets. In future work we will first compare the results of different home definition methods against available ground truth for different types of datasets, then classify datasets into different categories and finally give specific recommendations of which method to use for datasets belonging to a particular category.

\section*{Acknowledgment}
The authors would like to thank Banco Bilbao Vizcaya Argentaria (BBVA) for providing the dataset for this research. Special thanks to Juan Murillo Arias, Marco Bressan, Elena Alfaro Martinez, Mar\'ia Hern\'andez Rubio and Assaf Biderman for organizational support of the project and stimulating discussions. We further thank BBVA, MIT SMART Program, Center for Complex Engineering Systems (CCES) at KACST and MIT, Accenture, Air Liquide, The Coca Cola Company, Emirates Integrated Telecommunications Company, The ENELfoundation, Ericsson, Expo 2015, Ferrovial, Liberty Mutual, The Regional Municipality of Wood Buffalo, Volkswagen Electronics Research Lab, UBER, and all the members of the MIT Senseable City Lab Consortium for supporting the research. Finally, the authors also acknowledge support of the research project "Managing Trust and Coordinating Interactions in Smart Networks of People, Machines and Organizations", funded by the Croatian Science Foundation.

\bibliographystyle{splncs03}
\bibliography{literature}

\end{document}